\documentclass[12pt, a4paper]{article}
\usepackage{cite}
\usepackage{amsmath,amssymb}
\input{colordvi.tex}
\usepackage[usenames,dvipsnames]{color}
\usepackage{graphicx}
\begin{document}

%%%%%%%%%%%%%%%%%%%%%%%%%%%%%%%%%%%%%%%%%%%%%%%%%%
\begin{titlepage}

\begin{center}

\hfill UT-16-38, TU-1038,  IPMU 16-0213\\

\vskip .75in

{\Large \bf 
Gravitational waves from domain walls \\and their implications
}

\vskip .75in

{\large Kazunori Nakayama$^{a,b}$, Fuminobu Takahashi$^{c,b}$ and Norimi Yokozaki$^c$}

\vskip 0.25in

\begin{tabular}{ll}
$^a$ &\!\! {\em Department of Physics, Faculty of Science, }\\
& {\em The University of Tokyo,  Bunkyo-ku, Tokyo 133-0033, Japan}\\[.3em]
$^{b}$ &\!\! {\em Kavli IPMU (WPI), UTIAS,}\\
&{\em The University of Tokyo,  Kashiwa, Chiba 277-8583, Japan}\\[.3em]
$^{c}$ &\!\! {\em Department of Physics, Tohoku University, }\\
& {\em Sendai, Miyagi 980-8578, Japan}

\end{tabular}

\end{center}
\vskip .5in

\begin{abstract}
We evaluate the impact of domain-wall annihilation on the
currently ongoing and planned gravitational wave experiments,
including a case in which domain walls experience a frictional 
force due to interactions with the ambient plasma. We show the sensitivity reach
in terms of physical parameters, namely, the wall tension and the
annihilation temperature. 
We find that a Higgs portal scalar, which stabilizes the Higgs potential 
at high energy scales, can form domain walls whose annihilation 
produces a large amount of gravitational waves within the reach
of the advanced LIGO experiment (O5). Domain wall annihilation
can also generate baryon asymmetry if the scalar is coupled to either
SU(2)$_L$ gauge fields or the $(B-L)$ current. This is a variant of spontaneous
baryogenesis, but it naturally avoids the isocurvature constraint
due to the scaling behavior of the domain-wall evolution. 
We delineate the parameter space where
the domain-wall baryogenesis works successfully and discuss 
its implications for the gravitational wave experiments.
\end{abstract}

\end{titlepage}

%\tableofcontents

\renewcommand{\thepage}{\arabic{page}}
\setcounter{page}{1}
\renewcommand{\thefootnote}{\#\arabic{footnote}}
\setcounter{footnote}{0}
%%%%%%%%%%%%%%%%%%%%%%%%%%%%%%%%%%%%%%%%%%%%%%%%%%

\newpage

%%%%%%%%%%%%%%%%%%%%%%%%%%%%%%%%%%%%%%%%%%%%%
\section{Introduction} 
%%%%%%%%%%%%%%%%%%%%%%%%%%%%%%%%%%%%%%%%%%%%%
Symmetry and its breaking have been a guiding principle in modern physics. 
Some symmetry  restored at high temperatures may be eventually 
spontaneously broken at low temperatures. If the breaking scale is high, 
such spontaneously broken symmetry is hidden and inaccessible to direct observation, but
 the phase transition might have left some traces, i.e., topological 
defects such as monopoles, cosmic strings, and domain 
walls~\cite{Zeldovich:1974uw,Kibble:1976sj,Vilenkin:1981zs}. 
Thus, topological defects can be a probe of high-energy physics, and they 
could play an important role in cosmology. 

%Topological defects 
%If the original symmetry is exact,  the corresponding topological defects are stable. On the other hand, if it is
%approximate, the topological defects are unstable and decay with a finite lifetime. 
If the discrete symmetry is spontaneously broken, domain walls are formed.
Then, in a few Hubble times, domain walls start to follow a scaling law, i.e., the averaged number of the
walls per a Hubble horizon remains constant over 
time~\cite{Press:1989yh,Garagounis:2002kt,Oliveira:2004he,Leite:2011sc}. 
As a result, their energy density
%Domain walls are cosmologically important, because its 
%The energy density of domain walls 
decreases more slowly than radiation or matter, and so,  
 domain walls would easily dominate the Universe and spoil the success of 
the standard cosmology. To avoid the cosmological catastrophe, domain walls must either be 
unstable or remain subdominant until present. 

%The domain-wall cosmology has been studied extensively in the literature (see e.g. \cite{Vilenkin:2000jqa}).
If the discrete symmetry is approximate, domain walls are unstable. There is an energy bias between the true vacuum
and the false vacua, which induces a pressure on the walls.  Then, 
 domain walls annihilate when the pressure on the walls 
becomes larger than the tension of domain walls. Such domain-wall annihilation is so violent that
it leads to emission of a large amount of gravitational waves (GWs).
Thus produced GWs retain various information on the UV physics. For instance,
gaugino condensation leads to the formation of domain walls as a result of the spontaneous
break down of the discrete R symmetry~\cite{Dvali:1996xe}, which is further explicitly broken by a constant term in 
the superpotential. The produced GWs are expected to be peaked at a frequency
determined by the gravitino mass~\cite{Takahashi:2008mu,Dine:2010eb}. Domain walls and the associated
production of GWs are also studied in the standard model~\cite{Kitajima:2015nla} and 
its extension such as the Higgs portal model~\cite{Jaeckel:2016jlh}, 
the next-to-minimal supersymmetric Standard Model~\cite{Hamaguchi:2011nm,Kadota:2015dza}, 
thermal inflation model~\cite{Moroi:2011be}
and the clockwork QCD axion~\cite{Higaki:2016yqk,Higaki:2016jjh}, etc.

The domain-wall annihilation also generates  baryon asymmetry, if the 
corresponding scalar field is coupled to either SU(2)$_L$ gauge fields or the 
$(B-L)$ current~\cite{Daido:2015gqa}. This is a variant of the so called spontaneous
baryogenesis~\cite{Cohen:1987vi,Dine:1990fj,Cohen:1991iu,Chiba:2003vp,isoc4,
Kusenko:2014uta,Ibe:2015nfa,DeSimone:2016ofp,DeSimone:2016juo}.
In contrast to  spontaneous baryogenesis scenarios
in the slow-roll regime, the baryonic isocurvature perturbation is naturally suppressed by
the scaling behavior of the domain-wall network.\footnote{See 
Refs.~\cite{isoc4, Chiba:2003vp,DeSimone:2016juo}
for other ways to avoid the isocurvature constraint.}

In this paper we focus on the  annihilation of unstable domain walls, and evaluate its 
impact on the currently ongoing and planned GW experiments. One of the purposes of
this paper is to delineate the parameter space in terms of physical parameters 
where the GWs produced by the domain-wall annihilation are within the reach of 
the present and future experiments. We also study the domain-wall baryogenesis and its associated 
GW emission. 

The rest of this paper is organized as follows. 
In Section 2, we evaluate the frequency and energy density of GWs from annihilating domain walls, 
and show the sensitivity reach of various GW experiments
 in terms of the domain wall tension and the annihilation temperature.
  In  Section 3, we discuss implications of the detection of the GWs. 
The section 4 is devoted to  conclusions.

%%%%%%%%%%%%%%%%%%%%%%%%%%%%%%%%%%%%%%%%%%%%%
\section{Gravitational waves from domain walls} 
%%%%%%%%%%%%%%%%%%%%%%%%%%%%%%%%%%%%%%%%%%%%%

%%%%%%%%%%%%%%%%%%%%%%%%%%%%%%%%%%%%%%%%%%%%%
\subsection{Biased domain walls} 
%%%%%%%%%%%%%%%%%%%%%%%%%%%%%%%%%%%%%%%%%%%%%

Let us consider a real scalar $\phi$ with a $Z_2$-symmetric double-well potential:
\begin{align}
	V(\phi) = \frac{\lambda_\phi}{4}(\phi^2-v_\phi^2)^2.
	\label{V}
\end{align}
%%
%where $V_\epsilon$ denotes the (small) $Z_2$ breaking term, which is called bias.
%First neglecting the $Z_2$ breaking term, 
One can  find a static solution of the equation of motion with a boundary
condition such that the two vacua are realized at $x \to \pm \infty$,
\begin{align}
	\phi({\bf x}) = v_\phi\tanh\left( \sqrt{\frac{\lambda_\phi}{2}} v_\phi x \right),
\end{align}
which represents a domain wall extending along the $x = 0$ plane.
The typical width of the domain wall $\delta$ is given by the inverse of the mass of $\phi$ 
at the potential minimum, 
$\delta \sim m_\phi^{-1} = (\sqrt{2\lambda_\phi}v_\phi)^{-1}$.
The surface density, i.e. the tension of the domain wall, is given by
\begin{align}
	\sigma = \int_{-\infty}^{\infty} dx \,\rho_\phi = \frac{2\sqrt 2}{3}\sqrt{\lambda_\phi} v_\phi^3 = \frac{2}{3}m_\phi v_\phi^2,
\end{align}
where $\rho_\phi = \frac{1}{2} |\nabla \phi|^2 + V(\phi)$ is the (static) energy density of $\phi$.

When the $Z_2$ symmetry is spontaneously broken in the early Universe,
domain walls are formed, and they quickly follow a scaling solution such that
%Once the domain wall is formed in a cosmological evolution by e.g. the phase transition, 
%its energy density tends to soon dominate the Universe.
%Since the domain wall reaches the scaling solution in which
 there is about one domain wall per Hubble horizon~\cite{Press:1989yh,Garagounis:2002kt,Oliveira:2004he,Leite:2011sc}. Then the energy density of the domain 
 walls scales as $\rho_{\rm DW} \sim \sigma H$, and its fraction to the total energy density is 
 %given by
%%
\begin{align}
	r_{\rm DW}\equiv \frac{\rho_{\rm DW}}{\rho_{\rm tot}} \simeq \frac{\sigma H}{3H^2 M_P^2}
	% \sim \frac{\sigma}{HM_P^2}
	\sim \frac{H_{\rm dom}}{H},
\end{align}
where $H$ denotes the Hubble parameter and $M_P$ the reduced Planck scale and we have defined
\begin{align}
	H_{\rm dom} \equiv \frac{\sigma}{M_P^2}.  \label{Hdom}
\end{align}
Thus, domain walls start to dominate the Universe at $H \sim H_{\rm dom}$, and the Universe will be
extremely inhomogeneous afterwards.  This is the notorious cosmological domain wall problem.

To avoid the cosmological domain wall problem, let us introduce a small $Z_2$ breaking term, which induces an energy  bias 
between the two vacua and thereby destabilize the walls.
%which gives a preference for one of the two vacua and destabilizes the domain walls.
The difference in the energy density between two vacua, $V_\epsilon$, 
 gives a negative pressure in domains with the false vacuum and the walls are accelerated by this pressure.
As a result the domains with the false vacuum collapse at
\begin{align}
	H_{\rm ann} \sim \frac{V_\epsilon}{\sigma}.  \label{Hdec}
\end{align}
The energy fraction of domain walls at the annihilation is given by
\begin{align}
	r_{\rm DW}(H_{\rm ann}) \sim \frac{\sigma^2}{V_\epsilon M_P^2} \sim \frac{H_{\rm dom}}{H_{\rm ann}}.
\end{align}
Thus, domain walls annihilate before they start to dominate the Universe if $V_\epsilon > \sigma^2/M_P^2$.

%%%%%%%%%%%%%%%%%%%%%%%%%%%%%%%%%%%%%%%%%%%%%
\subsection{Gravitational waves from annihilating domain walls} 
%%%%%%%%%%%%%%%%%%%%%%%%%%%%%%%%%%%%%%%%%%%%%

The domain-wall annihilation is an efficient source of GWs.
The power emitted as GWs from a massive object 
%with a quadrupole moment $I$
is simply estimated by the quadrupole radiation formula:
\begin{align}
	\dot E_{\rm GW} \sim \frac{1}{M_P^2}\left( \frac{d^3 I}{dt^3} \right)^2,  \label{quad}
\end{align}
where $I$ is a quadrupole moment of the object.
In the scaling regime, there is one or a few domain walls in one Hubble horizon.
For such a domain wall stretched to the horizon scale, the quadrupole moment is estimated to
be  $I \sim \sigma / H^4$.
Thus the energy density of GWs emitted during one Hubble time is given by
\begin{align}
	\rho_{\rm GW} \sim \frac{\dot E_{\rm GW} H^{-1}}{H^{-3}} \sim \frac{\sigma^2}{M_P^2}.
\end{align}
The present GW energy density emitted at $H = H_e\, ( > H_{\rm ann}$) normalized by the critical density today is given by
\begin{align}
	\Omega_{\rm GW} \simeq \Omega_{\rm r} \,\beta
	 \frac{\sigma^2}{3H_e^2 M_P^4}  \sim \Omega_{\rm r} \,\beta \left(r_{\rm DW}(H_e)\right)^2
	 \label{OGW}
\end{align}
where $\Omega_{\rm r} \simeq 8.5\times 10^{-5}$ denotes the density parameter of radiation (i.e. photons and three massless neutrinos) 
and we have defined
\begin{align}
	\beta \equiv \left( \frac{g_*(H_e)}{g_{*0}} \right)  \left( \frac{g_{*s0}}{g_{*s}(H_e)} \right)^{4/3} \simeq 0.39,
\end{align}
where we have substituted  $g_{*s0} \simeq 3.909$, $g_{*0}\simeq 3.363$, 
 $g_*(H_e) = g_{s*}(H_e) = 106.75$ in the last equality.
Here and in what follows we assume that the Universe is radiation dominated during the epoch of GW emission.
The typical frequency of GW at the emission is of order $H_e$, and the corresponding frequency at present is given by
\begin{align}
	f (H_e)= \frac{H_e}{2\pi} \frac{a(H_e)}{a_0} \simeq 3\,{\rm Hz} \left( \frac{g_*(H_e)}{106.75} \right)^{1/6}\left( \frac{T_e}{10^8\,{\rm GeV}} \right),
\end{align}
where $T_e$ denotes the temperature at $H=H_e$.
Therefore, Eq.~(\ref{OGW}) indicates that the present GW spectrum has a peak at the frequency $f_{\rm peak}=f(H_{\rm ann})$.
The peak GW energy density is given by
\begin{align}
	\Omega_{\rm GW, peak} \simeq \Omega_{\rm r}\beta \left( r_{\rm DW}(H_{\rm ann}) \right)^2.
\end{align}
Moreover, according to the numerical simulation~\cite{Hiramatsu:2013qaa}, the 
high and low frequency tails of $\Omega_{\rm GW}$ scale as $f^{-1}$ and $f^{3}$, respectively.
Thus the whole GW spectrum is approximated by
\begin{align}
	\Omega_{\rm GW} \simeq \Omega_{\rm GW, peak}\times 
	\begin{cases}
		\displaystyle{\left( \frac{f_{\rm peak}}{f} \right)} & {\rm for}~~f>f_{\rm peak}\\
		&\\
		\displaystyle{\left( \frac{f}{f_{\rm peak}} \right)^3 }& {\rm for}~~f<f_{\rm peak}
	\end{cases}.
\end{align}
Note that numerical simulation~\cite{Hiramatsu:2013qaa} also implies that there is a high-frequency cutoff 
at $f > f_{\rm cut}$, corresponding to the width of domain-wall at the annihilation:
$f_{\rm cut} \sim f_{\rm peak} (m_\phi/H_{\rm ann})$.

The system of biased domain wall is conveniently parameterized by the domain wall tension $\sigma$ and the 
temperature at which domain walls annihilate, $T_{\rm ann} \sim \sqrt{H_{\rm ann} M_P}$.
There are several conditions on these parameters.
First, domain walls must annihilate before they dominate the Universe: $r_{\rm DW}(H_{\rm ann})\lesssim 1$.
This leads to
\begin{align} 
	\sigma \lesssim T_{\rm ann}^2 M_P.  \label{Tann_const1}
\end{align}
On the other hand, we assume that the bias energy is small enough to be regarded as a perturbation to the original scalar potential:
$V_\epsilon \lesssim \lambda_\phi v_\phi^4$. It is translated as
\begin{align}
	T_{\rm ann}^2 \lesssim m_\phi M_P \sim  (\lambda_\phi \sigma)^{1/3} M_P.  \label{Tann_const2}
\end{align}
The above constraints can be expressed as
\begin{align}
	%1 \gtrsim r_{\rm DW}(H_{\rm ann}) \gtrsim \left(\frac{v_\phi}{M_P}\right)^{2}.
	 \left(\frac{v_\phi}{M_P}\right)^{2} \lesssim r_{\rm DW}(H_{\rm ann}) \lesssim 1.
	 \label{const}
\end{align}
%%
%Thus for larger GW amplitude (larger $r_{\rm DW}(H_{\rm ann})$), the peak frequency becomes smaller.
The GW amplitude takes the largest value if the domain walls annihilate just before they start
to dominate the Universe, i.e, $r_{\rm DW}(H_{\rm ann}) = {\cal O}(0.1)$.
For example, let us consider the following interaction as a possible $Z_2$ breaking term,
\begin{align}
	\Delta V = \frac{1}{2} \mu^3 \phi,
\end{align}
which induces a bias energy $V_\epsilon \simeq \mu^3 v_\phi$ between the two vacua.
%In order for this term to be regarded as a perturbation to the original potential, $\mu \lesssim \lambda_\phi^{1/3} v_\phi$ should be satisfied.
%Combined with the constraint $r_{\rm DW}(H_{\rm ann}) < 1$, we need
The  constraint (\ref{const}) is written as
\begin{align}
%	v_\phi \left( \frac{m_\phi}{M_P} \right)^{2/3} \lesssim \mu \lesssim \left(\frac{m_\phi^{5/3}v_\phi^{7/3}}{M_P}\right)^{1/3}.
v_\phi \left( \frac{m_\phi}{M_P} \right)^{2/3} \lesssim \mu \lesssim  \lambda_\phi^{1/3} v_\phi.
\end{align}

If the scalar $\phi$ is in thermal contact with the ambient plasma, 
the scalar potential may receive thermal corrections. Such
 thermal effects on the $\phi$ potential is negligible at $T=T_{\rm ann}$, if
\begin{align}
	T_{\rm ann} \lesssim v_\phi \sim \lambda_\phi^{-1/6} \sigma^{1/3}.  \label{Tann_const3}
\end{align}
This is almost always stronger than (\ref{Tann_const2}) unless $\lambda_\phi$ is extremely small. 
We note however that, if $\phi$ has interactions with thermal plasma, the domain-wall
evolution can be significantly affected, as we shall see in the next subsection.
%Thus, combining (\ref{Tann_const1}) and (\ref{Tann_const2}), we obtain
%%%
%\begin{align}
%	\left(\frac{\sigma}{M_P}\right)^{1/2} \lesssim T_{\rm ann} \lesssim \sigma^{1/3}.  \label{Tann_const4}
%\end{align}
%%%
%The above constraints can also be expressed as
%%%
%\begin{align}
%	%1 \gtrsim r_{\rm DW}(H_{\rm ann}) \gtrsim \left(\frac{v_\phi}{M_P}\right)^{2}.
%	 %\left(\frac{v_\phi}{M_P}\right)^{2} \lesssim r_{\rm DW}(H_{\rm ann}) \lesssim 1.
%	 \frac{\sigma^{1/3}}{M_P} \lesssim r_{\rm DW}(H_{\rm ann}) \lesssim 1.
%	 \label{const}
%\end{align}
%%%

%For example, let us consider the following interaction as a possible $Z_2$ breaking term,
%%%
%\begin{align}
%	\Delta V = \frac{1}{2} \mu^3 \phi,
%\end{align}
%%%
%which induces a bias energy $V_\epsilon \simeq \mu^3 v_\phi$ between the two vacua.
%%In order for this term to be regarded as a perturbation to the original potential, $\mu \lesssim \lambda_\phi^{1/3} v_\phi$ should be satisfied.
%%Combined with the constraint $r_{\rm DW}(H_{\rm ann}) < 1$, we need
%The  constraint (\ref{const}) is written as
%%%
%\begin{align}
%	v_\phi \left( \frac{m_\phi}{M_P} \right)^{2/3} \lesssim \mu \lesssim \left(\frac{m_\phi^{5/3}v_\phi^{7/3}}{M_P}\right)^{1/3}.
%\end{align}
%%%

%%%%%%%%%%%%%%%
\begin{figure}[!t]
\begin{center}
\includegraphics[scale=1.3]{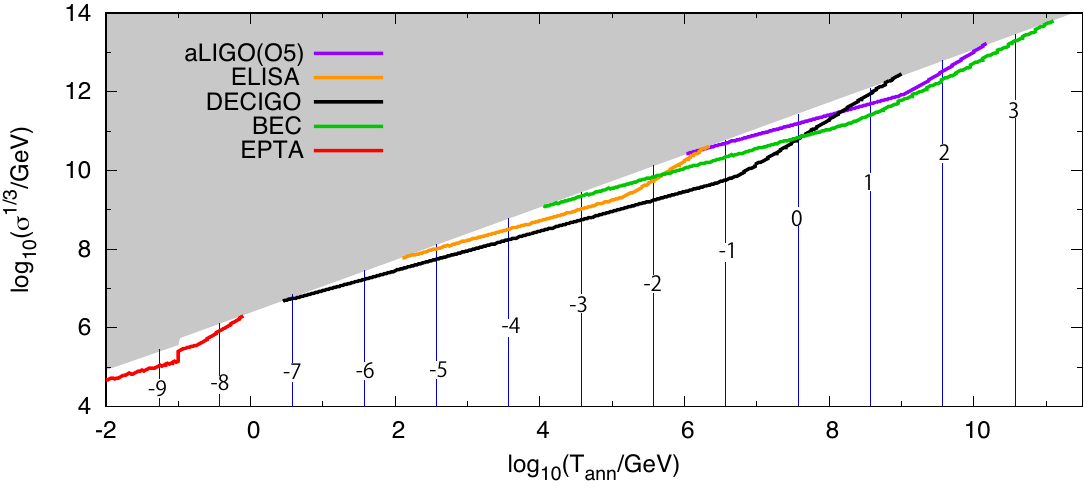}
%\includegraphics[scale=1.3]{fig2a.pdf}
%\includegraphics[scale=1.3]{fig3a.pdf}
%\includegraphics[scale=1.3]{fig4a.pdf}
%\qquad
%\includegraphics[width=.40\textwidth, bb=0 0 178 188]{gcu_b.pdf}
\caption{
The sensitivities of the GWs on $T_{\rm ann}$\,-\,$\sigma^{1/3}$ plane. 
The blue solid line shows $\log_{10} (f_{\rm peak}/[{\rm Hz}])$.
}
\label{fig:dw_and_gw1}
\end{center}
\end{figure}
%%%%%%%%%

%%%%%%%%%%%%%%%
\begin{figure}[!t]
\begin{center}
\includegraphics[scale=1.3]{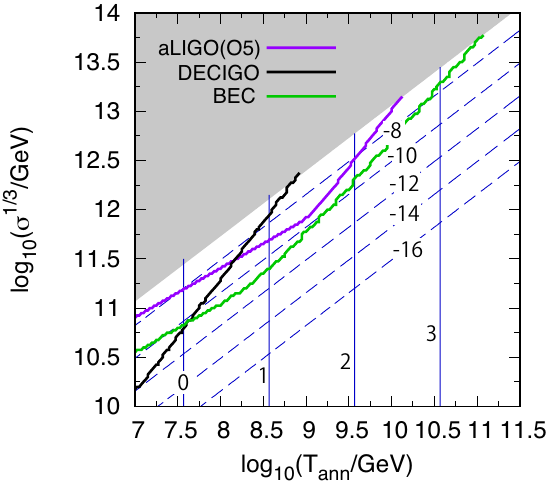}
\includegraphics[scale=1.3]{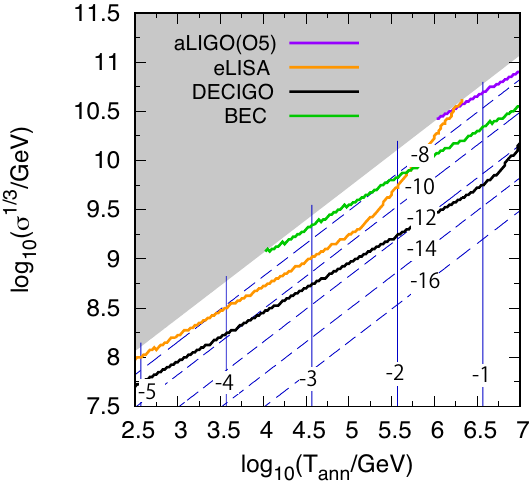}
\includegraphics[scale=1.3]{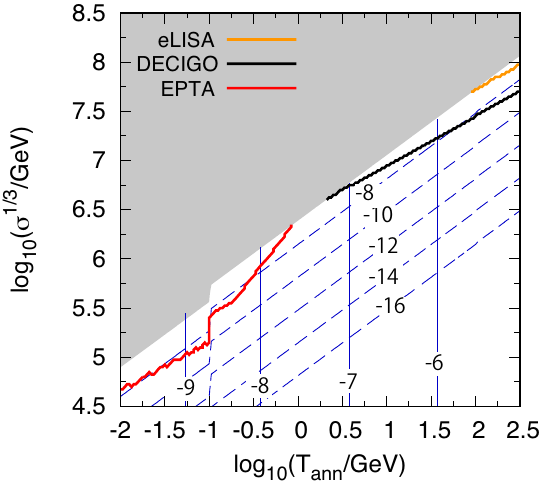}
%\qquad
%\includegraphics[width=.40\textwidth, bb=0 0 178 188]{gcu_b.pdf}
\caption{
Zooms of Fig.~\ref{fig:dw_and_gw1}.
The blue solid and dashed lines show $\log_{10} (f_{\rm peak}/[{\rm Hz}])$ and $\log_{10} (\Omega_{\rm GW, peak} h^2)$, respectively.
%The sensitivities of the gravitational waves on $T_{\rm ann}$\,-\,$\sigma^{1/3}$ plane. 
%The blue solid and dashed lines show $\log_{10} (f_{\rm peak})$ and $\log_{10} (\Omega_{\rm GW, peak} h^2)$, respectively.
}
\label{fig:dw_and_gw2}
\end{center}
\end{figure}
%%%%%%%%%

\vspace{10pt}
The produced GWs can be detected by various experiments such as advanced LIGO (aLIGO)~\cite{aligo}, DECIGO~\cite{decigo}, eLISA~\cite{elisa}, and EPTA~\cite{epta}. Also, there is a recently proposed experiment using a Bose-Einstein condensate (BEC)~\cite{Sabin:2015mha}. 
In Fig.~\ref{fig:dw_and_gw1}, we show the sensitivity reach of those experiments on the $T_{\rm ann}$-$\sigma^{1/3}$ plane. In the regions above the purple, orange, black, green and red solid curves,
the produced GWs are within the reach of aLIGO (O5), eLISA, DECIGO, BEC and EPTA, respectively.  
In Fig.~\ref{fig:dw_and_gw2}, zooms of Fig.~\ref{fig:dw_and_gw1} are presented.
We also show contours of $\log_{10} (f_{\rm peak}/[{\rm Hz}])$ (vertical blue solid lines) and $\log_{10} (\Omega_{\rm GW, peak} h^2)$ (blue dashed lines). 
Here, $h \approx 0.68$~\cite{pdg} is a scale factor for Hubble expansion rate. 
The top left triangle region shaded in the gray is excluded since the domain walls dominate the Universe before they annihilate.
%In the bottom right gray region, $T_{\rm ann}$ is too large and the symmetry would be restored at $T=T_{\rm ann}$.
%The region between these two grays satisfy the constraint (\ref{Tann_const4}).
%In the bottom right triangle region, $T_{\rm ann} > \sigma^{1/3}$.

%%%%%%%%%%%%%%%%%%%%%%%%%%%%%%%%%%%%%%%%%%%%%
\subsection{Effects of friction} 
%%%%%%%%%%%%%%%%%%%%%%%%%%%%%%%%%%%%%%%%%%%%%

So far we have ignored possible domain-wall interactions with the environmental plasma.
In some realistic setup, e.g. the Higgs portal case to be discussed in the next section, the domain wall interaction with thermal plasma can be significant.
It causes a frictional force on domain walls and can change the domain wall evolution in the early epoch.

Let us suppose that a $\chi$ particle, which interacts with $\phi$, is reflected around the domain wall with 
a probability close to be unity.
If $\chi$ is in thermal equilibrium, the frictional force per unit area of domain wall is given by~\cite{Vilenkin:2000jqa}
\begin{align}
	F_\chi \sim n_\chi \Delta p \sim T^4 v,
\end{align}
where $n_\chi \sim T^3$ is the $\chi$ number density, $\Delta p \sim Tv$ the average momentum transfer per collision,
with $v$ being the domain wall velocity.
On the other hand, the domain wall tension itself causes accelerating force:
\begin{align}
	F_\sigma \sim \frac{\sigma}{R},
\end{align}
where $R$ is the curvature radius of the domain wall.
Note that just after the phase transition, $R$ is much smaller than the Hubble length and maybe rather close to $T^{-1}$.
After the time $t$, the irregularities at scales below $\sim vt$ is smoothed out, hence we expect $R\sim vt$.
If these two forces are balanced, we have
\begin{align}
	v \sim \left( \frac{H_{\rm dom}}{H} \right)^{1/2},~~~~~~~R\sim H^{-1}\left( \frac{H_{\rm dom}}{H} \right)^{1/2},   \label{vR_fric}
\end{align}
where $H_{\rm dom}$ is given in (\ref{Hdom}).
The energy density of domain wall is given by $\rho_{\rm DW}\sim \sigma/R$. Hence its energy fraction is estimated as
\begin{align}
	\tilde r_{\rm DW} = \frac{\rho_{\rm DW}}{\rho_{\rm tot}} \sim \left( \frac{H_{\rm dom}}{H} \right)^{1/2},
\end{align}
%%
%{\bf Hence, the domain wall dominates the Universe at $H \sim H_{\rm dom}$,
%and after that the domain wall enters in the scaling regime as in the case of without friction. }
where a tilde is added to show that it is evaluated in the friction regime.
One should note that the scaling at $H > H_{\rm dom}$ is significantly different.
The energy density of the domain wall is much higher and the velocity/curvature radius is much smaller than the case without friction.

The domain walls in the friction regime may annihilate when the bias energy becomes significant.
The collapsing force per unit area of the domain wall caused by the bias is estimated as
\begin{align}
	F_\epsilon \sim V_\epsilon.
\end{align}
Thus the bias-induced force begins to dominate over $F_\sigma (\sim F_\chi)$ at $H\sim \widetilde H_{\rm ann}$ with
\begin{align}
	\widetilde H_{\rm ann} \sim \left(\frac{V_\epsilon^{2}}{\sigma M_P^2}\right)^{1/3}\sim H_{\rm ann}^{2/3} H_{\rm dom}^{1/3},
\end{align}
where we again added a tilde as $\widetilde H_{\rm ann}$ to represent the annihilation time in the friction regime
and distinguish it from the annihilation epoch without friction, $H_{\rm ann}$, defined in (\ref{Hdec}).
Note that $H_{\rm ann} > \widetilde H_{\rm ann} > H_{\rm dom}$.
After that, the false vacuum region begins to collapse while the friction acts as a repulsive force against the collapse.
If they are balanced, i.e., $F_\epsilon \sim F_\chi$, we have $v\sim V_\epsilon/T^4$ hence it is accelerated as time goes on.
Actually we find that the typical curvature radius is $H_{\rm ann}^{-1}$ 
and the domain wall velocity $v\sim  \widetilde H_{\rm ann} /H_{\rm ann}$ at $H \sim  \widetilde H_{\rm ann} $.
Therefore, the typical time required for collapse of the false vacuum region is
\begin{align}
	\left(\frac{R}{v}\right)_{H\sim \widetilde H_{\rm ann} } \sim  \widetilde H_{\rm ann} ^{-1}.
\end{align}
It means that the false vacuum collapses within one Hubble time around $H\sim  \widetilde H_{\rm ann} $.
Thus domain walls annihilate when $H=\widetilde H_{\rm ann} $ in the friction-dominated case.
The energy fraction of the domain wall at the annihilation is given by
\begin{align}
	\tilde r_{\rm DW}( \widetilde H_{\rm ann} ) \sim \left( \frac{H_{\rm dom}}{ \widetilde H_{\rm ann}} \right)^{1/2}.
\end{align}

Now let us estimate GWs from the collapsing domain walls in the friction regime.
From the quadrupole formula (\ref{quad}), by substituting $I \sim \sigma R^4$, we have
\begin{align}
	\dot E_{\rm GW} \sim v^6 \left(\frac{\sigma R}{M_P}\right)^2,
\end{align}
as a GW power emitted from the domain wall region with typical curvature radius $R$.
There are $(RH)^{-3}$ such regions in the Hubble volume, and hence the averaged GW energy density emitted during one Hubble time is
estimated as
\begin{align}
	\rho_{\rm GW}(H) \sim \frac{\dot E_{\rm GW} H^{-1}}{R^3} \sim \frac{v^6}{HR} \left( \frac{\sigma}{M_P} \right)^{2}. 
\end{align}
Substituting (\ref{vR_fric}), we find
\begin{align}
	r_{\rm GW}(H) = \frac{\rho_{\rm GW}}{\rho_{\rm tot}} \sim \left( \frac{H_{\rm dom}}{H} \right)^{9/2} \sim \left(\tilde r_{\rm DW}(H)\right)^9.
\end{align}
Thus the peak GW energy density at $f_{\rm peak} = f( \widetilde H_{\rm ann})$ is 
\begin{align}
	\Omega_{\rm GW, peak} \simeq \Omega_{\rm r}\beta \left( \tilde r_{\rm DW}( \widetilde H_{\rm ann}) \right)^9.
\end{align}
We see that the energy density of the GWs is smaller than the case without a friction.

\begin{figure}[!t]
\begin{center}
\includegraphics[scale=1.3]{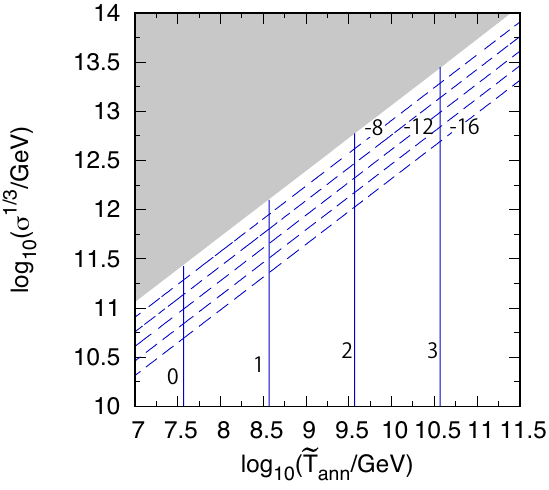}
\includegraphics[scale=1.3]{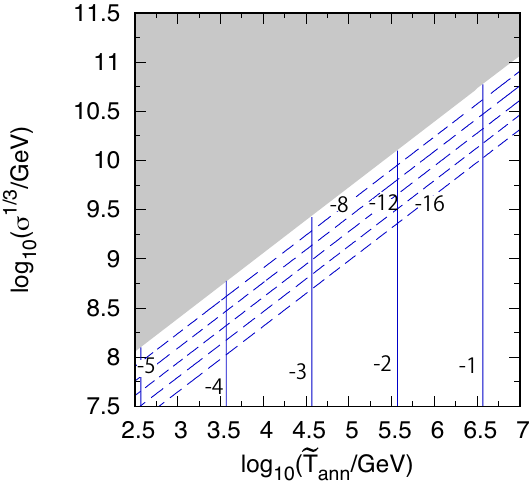}
\includegraphics[scale=1.3]{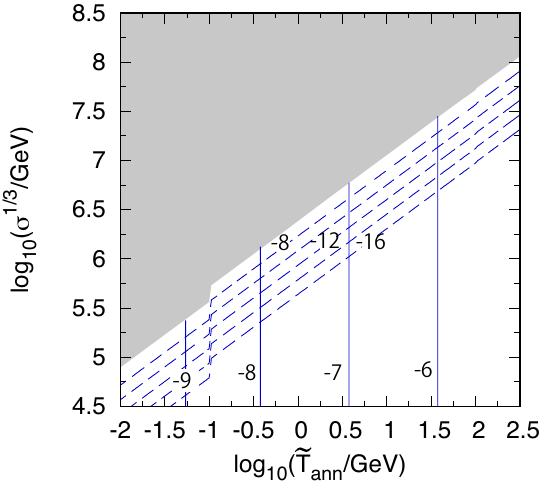}
%\qquad
%\includegraphics[width=.40\textwidth, bb=0 0 178 188]{gcu_b.pdf}
\caption{
The contours of $f_{\rm peak}$ and $\Omega_{\rm GW, peak} h^2$ with the frictional force.
The blue solid and dashed lines show $\log_{10} (f_{\rm peak}/[{\rm Hz}])$ and $\log_{10} (\Omega_{\rm GW, peak} h^2)$, respectively.
%The sensitivities of the gravitational waves on $T_{\rm ann}$\,-\,$\sigma^{1/3}$ plane. 
%The blue solid and dashed lines show $\log_{10} (f_{\rm peak})$ and $\log_{10} (\Omega_{\rm GW, peak} h^2)$, respectively.
}
\label{fig:dw_and_gw3}
\end{center}
\end{figure}

\vspace{10pt}
In Fig.~\ref{fig:dw_and_gw3}, we show the contours of $f_{\rm peak}$ and $\Omega_{\rm GW, peak} h^2$ including effects of the friction. We denote the temperature at $H = \tilde H_{\rm ann}$ as $\tilde{T}_{\rm ann}$. 
It can be seen that, 
for a fixed $\tilde T_{\rm ann}$, the energy density of the GWs reduces more rapidly than the case without the friction 
as $\sigma$ becomes small. 
However, in the regions of $\tilde H_{\rm ann} \sim H_{\rm dom}$, 
$\Omega_{\rm GW, peak} h^2$ is almost same as the case without the friction (see Fig.~\ref{fig:dw_and_gw2}); therefore, these regions are expected to be covered by the various GW experiments.

%%%%%%%%%%%%%%%
%%%%%%%%%

%%%%%%%%%%%%%%%%%%%%%%%%%%%%%%%%%%%%%%%%%%%%%
\section{Phenomenological and cosmological implications} 
%%%%%%%%%%%%%%%%%%%%%%%%%%%%%%%%%%%%%%%%%%%%%

%%%%%%%%%%%%%%%%%%%%%%%%%%%%%%%%%%%%%%%%%%%%%
\subsection{Vacuum stability} 
%%%%%%%%%%%%%%%%%%%%%%%%%%%%%%%%%%%%%%%%%%%%%

It is known that the measured Higgs mass at the LHC implies that our present electroweak vacuum is metastable,
since the Higgs four-point coupling becomes negative above an intermediate scale $\Lambda$ due to the renormalization group running effect.
The instability scale $\Lambda$ is sensitive to the top quark mass and it is estimated to be $\Lambda\sim 10^{10}-10^{12}$\,GeV. Such a metastable vacuum can cause a cosmological problem
if the reheating temperature is high or if the Higgs field is to be identified with the inflaton~\cite{Bezrukov:2007ep}.

A real singlet scalar field with a large vacuum expectation value is motivated from a
viewpoint of the vacuum stability~\cite{Lebedev:2012zw,EliasMiro:2012ay}.
Let us consider a Higgs portal coupling of the singlet scalar as
\begin{align}
	V = \lambda_H\left( |H|^2 - v_{\rm EW}^2 \right)^2 + \frac{\lambda_\phi}{4}\left(\phi^2-v_\phi^2\right)^2 
	+ 2\lambda_{H\phi}\left( |H|^2 - v_{\rm EW}^2 \right)\left( \phi^2 - v_{\phi}^2\right).
\end{align}
The potential minimum is $\left<|H|^2\right> = v_{\rm EW}^2$ ($v_{\rm EW}=174$\,GeV) and 
$\left<\phi\right>=v_\phi$ if $\lambda_H \lambda_\phi > 4\lambda_{H\phi}^2$. 
Below the scale $m_\phi$, $\phi$ can be integrated out, which  results in the effective Higgs potential,
\begin{align}
	V_{\rm eff}(H) = \lambda_{\rm eff} (|H|^2-v_{\rm EW}^2)^2,
\end{align}
where
\begin{align}
	\lambda_{\rm eff} = \lambda_H - \frac{4\lambda_{H\phi}^2}{\lambda_\phi}.
\end{align}
Since this is the ``measured'' Higgs four-point coupling at the LHC, the genuine four-point coupling $\lambda_H$ differs from $\lambda_{\rm eff}$
by an amount of $4\lambda_{H\phi}^2/\lambda_\phi$ due to the threshold correction at the scale $m_\phi$.
Therefore, if $m_\phi$ is below the instability scale and $4\lambda_{H\phi}^2/\lambda_\phi$ is sizable 
we can have absolutely stable Higgs potential for $\lambda_{H\phi} >0$.

This is an interesting model, but it suffers from the domain wall problem unless the $Z_2$ symmetry is spontaneously broken during inflation and
never restored thereafter.\footnote{
	If the real scalar $\phi$ is replaced by a complex scalar with global/gauged U(1) symmetry, cosmic strings are formed instead of domain walls.
	The formation of cosmic strings is less harmful than the domain walls. However, if the U(1) symmetry is further explicitly broken to $Z_N$, our arguments can be applied to the corresponding domain walls. 
}
However, the Higgs-portal coupling tends to stabilize $\phi$ at the origin in the early Universe due to the finite-density correction and/or
the large Higgs field value in the Higgs inflation scenario.
In such a case, the domain walls are necessarily produced and we need to introduce a small $Z_2$ breaking bias term in the potential to destabilize the domain walls
as shown above.

The domain walls in this model experience a frictional force due to their interactions with the ambient plasma. 
In particular, the Higgs field takes a large field value inside the domain wall, which implies that most of the standard model particles (except for photons) become heavy inside the wall. Therefore, when those particles scatter with the domain wall, they are likely reflected from the wall, giving a sizable pressure on the wall. Thus, the domain wall in the Higgs portal scenario will be in the frictional regime except for the time before it starts to dominate the Universe. 
Note that we need $m_\phi = \sqrt{2\lambda_\phi}v_\phi \ll \Lambda$ to ensure the absolute stability of the Higgs potential. Therefore, $\sigma^{1/3}$ is smaller than the instability scale $\Lambda$ for $\lambda_\phi = {\cal O}(1)$.
One can see from Figs.~\ref{fig:dw_and_gw1}, \ref{fig:dw_and_gw2} and \ref{fig:dw_and_gw3}, the produced GWs can be within the reach of various GW experiments.

%%%%%%%%%%%%%%%%%%%%%%%%%%%%%%%%%%%%%%%%%%%%%
\subsection{Baryogenesis} 
%%%%%%%%%%%%%%%%%%%%%%%%%%%%%%%%%%%%%%%%%%%%%
Domain walls move around at a velocity close to the speed of light due to their tension.
If the corresponding scalar field is coupled to either SU(2)$_L$ gauge fields or to the $({B-L})$ 
current, the domain-wall dynamics induces an effective chemical potential for the baryon or lepton
number, and a right amount of baryon asymmetry could be generated when the domain walls 
annihilate~\cite{Daido:2015gqa}. 
This is a variant of the spontaneous baryogenesis scenario~\cite{Cohen:1987vi,Dine:1990fj,Cohen:1991iu}. 
In contrast to the spontaneous baryogenesis in the slow-roll regime, the scaling behavior
of the domain-wall network suppresses the baryonic isocurvature perturbations.

Here we estimate the amount of baryon asymmetry induced by domain walls and  delineate under 
which circumstances this scenario is valid. Let us assume that the scalar field $\phi$ is a pseudo scalar, 
and it is coupled to the lepton current as follows,
\begin{align}
{\cal L} \supset \frac{\partial_\mu \phi}{M} j^\mu_L,
\end{align}
where $M$ is a cut-off scale. Alternatively, we may introduce the coupling to the SU(2)$_L$ gauge fields like
\begin{align}
{\cal L} \supset \frac{\phi}{M} F^a_{\mu \nu} \tilde{F}^{a \mu \nu},
\end{align}
 and $F^a_{\mu \nu}$ is the field strength of SU(2)$_L$
gauge fields and $\tilde{F}^a_{\mu \nu}$ is its dual. In the latter case, the effective chemical potential is induced 
only when sphalerons are in equilibrium~\cite{Daido:2015gqa}. 
%{\bf As we shall see shortly, the resultant baryon
%asymmetry is independent of the cut-off scale, $M$.}

 In order to generate baryon asymmetry,
we need some operator which explicitly violates either baryon or lepton number. As a concrete example, 
we adopt the  $\Delta L = 2$ operators responsible for the neutrino masses in the seesaw 
mechanism~\cite{Minkowski:1977sc,Yanagida:1979as,Ramond1979,Glashow:1979nm}. Throughout this
paper we assume that right-handed neutrinos are so heavy that they are not produced in thermal 
plasma. The interaction rate for the $\Delta L = 2$ processes is approximated to be
\begin{align}
	\Gamma \sim \frac{T^3}{\pi^3} \frac{\sum m_i^2}{v_\mathrm{EW}^4},
	\label{dL=2}
\end{align}
where $m_i$ with $i=1,2,3$ denotes the mass of three active neutrinos~\cite{Buchmuller:2000nq}.
For the normal ordering for the neutrino mass differences, 
$\sum m_i^2 \simeq \Delta m^2_\mathrm{atm} \simeq 2.4 \times 10^{-3}~\mathrm{eV}^2$,
the decoupling temperature of the $L$-violating process is 
\begin{align}
\label{decT}
T_\mathrm{dec} \sim  3 \times 10^{13}\,\mathrm{GeV},
\end{align}
where we have assumed the radiation dominated Universe. As long as the reheating temperature $T_R$
is lower than $T_\mathrm{dec}$, the $L$-violating process is decoupled from the cosmic expansion.

Suppose that a domain wall passes through some point in space, where the effective chemical potential,
$\mu_{\rm eff} \sim \dot{\phi}/M \sim m_\phi v_\phi/M$, is induced temporarily. Then, some amount of lepton asymmetry
is induced because the chemical potential slightly biases $\Delta L = 2$ scattering processes,
$\ell \ell \leftrightarrow HH$, etc. In the scaling regime, however, domain walls move around
randomly and there is no preference for either of the vacua. As a result, no net asymmetry is generated.
On the other hand, when the domain walls decay away due to the energy bias between the vacua,
one of the sign of the chemical potential is preferred, leading to a non-zero asymmetry. The resultant
 baryon asymmetry is given by~\cite{Daido:2015gqa}
\begin{align}
\frac{n_B}{s} &\simeq \frac{28}{79}\times \frac{1}{2}\times \frac{45}{\pi^2 g_{*s}} \left.\frac{\Gamma}{T}\right|_{\rm ann},
\nonumber \\
&\simeq 2 \times 10^{-11} \left(\frac{T_{\rm ann}}{10^{11}{\rm GeV}}\right)^2,
\end{align}
where we have included the sphaleron conversion factor,  $g_{*s} = 106.75$ is the effective relativistic
degrees of freedom contributing to the entropy density, and we used the fact that the asymmetry is 
generated at the domain wall annihilation. 
Here, we take $M=v_{\phi}$ for simplicity.
In the second equality, we have substituted 
(\ref{dL=2}) assuming the radiation-dominated Universe. 

%Note that the final asymmetry depends only
%on the annihilation temperature, and it is independent of the domain-wall tension or the cut-off scale $M$.

Lastly let us comment on several conditions for the above scenario to work. First of all, the domain wall
must be sufficiently thick so that the dissipation of the asymmetry is neglected, i.e., $m_{\phi} < T_{\rm ann}$.
In terms of the tension of the domain wall, it reads
\begin{align}
\sigma & < \frac{T_{\rm ann}^3}{3 \lambda_\phi}.
\label{thick}
\end{align}
This condition is easily satisfied if $\lambda$ is much smaller than unity. For instance, instead of the
simple double-well potential (\ref{V}), we may consider a flat-top potential which is often considered
in a context of thermal inflation~\cite{Yamamoto:1985rd,Lyth:1995ka}. Secondly, the domain walls
could receive back reactions by the induced lepton asymmetry. This effect is negligible if the
energy stored in the domain walls is larger than the extra energy of the asymmetry:
\begin{align}
\rho_{\rm DW} & > \mu_{\rm eff} n_L,
\label{back}
\end{align}
where $n_L$ is the lepton number density. If this inequality is not satisfied, the domain walls would
feel the frictional force, and their evolution may be deviated from the simple scaling law. We impose
the condition (\ref{back}) at the time of domain wall annihilation, and then it reads
\begin{align}
\sigma & > 5 \times 10^{30}{\rm \,GeV}^3 \sqrt{\lambda_\phi}
 \left(\frac{T_{\rm ann}}{10^{11}{\rm \, GeV}} \right)^\frac{9}{2}.
\label{back2}
\end{align}
Thus, the domain-wall tension is bounded by the above two conditions (\ref{thick}) and (\ref{back2}),
and the allowed region will be enlarged for a smaller $\lambda_\phi$.

\begin{figure}[!t]
\begin{center}
\includegraphics[scale=1.0]{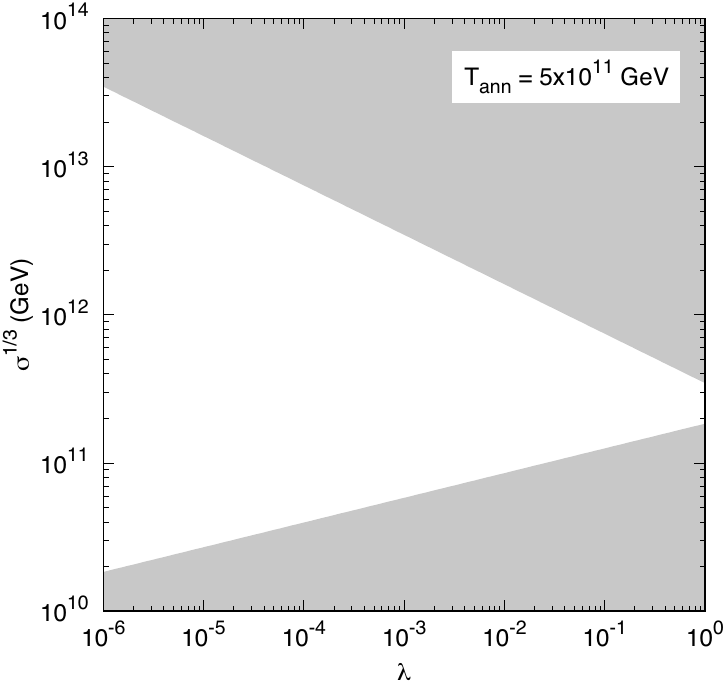}
%\qquad
%\includegraphics[width=.40\textwidth, bb=0 0 178 188]{gcu_b.pdf}
\caption{
The allowed region on $\lambda_\phi$-$\sigma^{1/3}$ plane for the successful baryogenesis. The gray regions are excluded.
}
\label{fig:bg}
\end{center}
\end{figure}

In Fig.~\ref{fig:bg} we have shown the parameter region where the domain-wall induced leptogenesis successfully 
explains the present baryon asymmetry of the Universe. 
%:
The gray regions are excluded since the constraint (\ref{thick}) and (\ref{back2}) are not satisfied. Comparing Fig.~\ref{fig:dw_and_gw1} with Fig.~\ref{fig:bg}, one can see that the GWs from the
domain wall annihilation are out of the reach of the aLIGO experiment due to the required $T_{\rm ann}$,
and they may be within the
projected sensitivity reach of the experiment using the Bose-Einstein condensate~\cite{Sabin:2015mha}.
Alternatively, if one considers another baryon- or lepton-number violating operator which is effective
down to temperatures as low as $T_{\rm ann} \sim 10^9$\,GeV, a similar baryogenesis could work at low temperatures
so that the GWs from the domain wall annihilation can be detected at the aLIGO experiment.

%%%%%%%%%%%%%%%
%%%%%%%%%

\section{Conclusions}

We have clarified the parameter regions of the domain wall tension and its annihilation temperature where the  produced GWs are detected by the ongoing and future experiments. 
If the Higgs potential is stabilized by a portal scalar, the mass of the portal scalar is likely to smaller than $\sim10^{12}$\,GeV. In this case the unstable domain wall may be generated and the tension of the domain wall is likely to be smaller than $\sim(10^{12}\,{\rm GeV})^3$. With this tension, the emitted GWs can be detected by aLIGO(O5), DECIGO, EPTA, and so on.
We have also discussed the possible effect of a frictional force originated from e.g., a Higgs portal interaction. In this case, the energy density of the GWs emitted from the domain wall annihilation is reduced compared to the case without the friction.

Finally, we show that the successful baryogenesis can be realized if the scalar forming the domain wall couples to the SU(2)$_L$ gauge fields or $(B-L)$ current. We considered a case in which the $\Delta = 2$ operator is responsible for
the generation of lepton number, and 
in this case, the produced GWs are out of reach of the aLIGO (O5). This is because the annihilation temperature is required to be rather high, and the  peaked frequency is higher than the sensitivity reach of the aLIGO.  On the other hand,  they may be detected by the GW experiment using BEC, which is likely to cover higher frequency range. Alternatively, if one considers another baryon/lepton number violating operator which is effective at temperatures of order $10^8$\,GeV, the GWs produced by the domain wall annihilation can be within the reach of the aLIGO.

\section*{Acknowledgments}

This work is supported by 
JSPS KAKENHI Grant Numbers JP15H05889 (F.T. and N.Y.), JP15K21733 (F.T. and N.Y.), 
JP26247042 (K.N. and F.T), JP15H05888 (K.N.), JP26287039 (F.T.), JP26800121 (K.N) and JP26104009 (K.N);
and by World Premier International Research Center Initiative (WPI Initiative), MEXT, Japan (K.N and F.T.).

%%%%%%%%%%%%%%%%%%%%%%%%%%%%%%%%%%%%%%%%%%%%%%%%%%

%%%%%%%%%%%%%%%%%%%%%%%%%%%%%%%%%%%%%%%%%%%%%%%%%%

\end{document}